# Fluctuations in the Statistical Model of Relativistic Heavy Ion Collisions

Aram Z. Mekjian
Department of Physics, Rutgers University, Piscataway, N.J. 08854

Abstract

Fluctuations in the statistical model of heavy ion collisions are studied. The role of statistics, relativity, constraints, decaying resonances and branching processes are investigated using this model. Also studied are thermodynamic properties related to these fluctuations with specific concerns centered around the behavior of the specific heat.

## I. Introduction

Event-by-event fluctuations in high energy collisions in small systems (e+,e-; e,p) have been of interest in the past and in large systems (heavy ion collisions) is gaining significant attention. The past interest centered around dynamical theories which would produce the negative binomial (NB) regularities and associated large non-poissonian fluctuations seen in experimental data. These NB regularities could be interpreted in terms of fractal behavior and intermittency associated with a possible underlying cascade process. Recent attention regarding the importance of fluctuations in the area of heavy ion collisions came first from the prediction of very large non-poissonian fluctuations in the neutral pion component coming from a disoriented chiral condensate [1]. More recently, attention has been drawn to the possibility of equilibrium charge fluctuations associated with the quark charges (2/3e, 1/3e) present in the quark-gluon [2-3]. Fluctuations also play an important role in phase transitions. Large density fluctuations appear at a critical point of a phase transition and give rise to the phenomena of critical opalescence. Large energy fluctuations are associated with a dramatic increase in the constant volume specific heat. Temperature fluctuations have been discussed in the context of heavy ion collision [4,5].

Fluctuations can have a purely statistical origin. Bose-Einstein correlations are present for pions and other mesons and Fermi-Dirac correlations are present for nucleons and other baryons. Such correlations have been used to study the space-time history of the collision using HBT techniques. Fluctuations are also influenced by constraints such as charge neutrality [6] and by decaying resonances as in a resonance gas [7]. Statistical models have been useful in describing one particle distributions [8-9]. Moreover, when statistical models are coupled with hydrodynamic expansion, transverse momentum spectra are accounted for [10]. Descriptions of the fluctuations in statistical models impose a more stringent test of their applicability. Some discussion of such fluctuations can be found in ref.[11]. Further extensions will be present here. Correlations also play an important role in understanding some features of multiparticle dynamics. Hierarchical correlation functions express all correlations in terms of two particle correlations [12]. Void scaling relations have been used in the past [13]. Clan variables [14,15] and combinants [16] have been introduced to characterize non-poissonian fluctuations. Various phenomenological models [17] based on quantum optics have been used to study multiplicity distributions. Wave packet descriptions [18] and generalized negative binomial models [19] are also used to study correlations. A general framework for studying correlations based on a grand canonical and canonical ensemble was also introduced in refs.[11,20,21].

In small systems, the multiplicity of produced particles is of the order of 10. For example the L3 experiment at Cern [22] has a mean multiplicity $<N>$=20.5. The H1 experiment Hera [23] at Desy has $<N>$=7.7. Associated fluctuations are large in both experiments. Specifically, $<N^2> - <N>^2 = <N> + B<N>^2$, B=.044 for L3 and .069 for H1 data. By contrast, RHIC experiments produce multiplicities in the thousands, With so many more particles, final state interactions over the space-time history of the system can dominate and lead to very different distributions dominated by Gaussian behavior with near poissonian fluctuations.

## 2. General Considerations

Multiplicity fluctuations can in general be written as a power series in the mean multiplicity $<N>$. In many cases they can be characterized by a simpler result that contains linear and quadratic terms in $<N>$, resulting in a form:

$$<N^2> - <N>^2 = A<N> + B<N>^2 \tag{1}$$

Here, A and B are constants. A somewhat more general result has been suggested in ref.[24] which is based on applications of random matrix theory to various physical phenomena. The more general result replaces $<N>^2$ with $<N>^\nu$, with $\nu$ a parameter. Values of $\nu$ different from 0 or 2 would be very interesting. An important example of eq.(1) is the Poisson distribution with A=1, B=0. Departures from this choice of A and B lead to non-poissonian fluctuations, and a classic example of this departure is the Planck distribution with A=1 and B=1. The Planck distribution is a special case of the negative binomial distribution where A=1, but B is arbitrary. The B=1/a with the constant "a" the traditional NB parameter. All the cases that will be considered here have $\nu$=2,0.

The Maxwell-Boltzmann limit in statistical physics results in fluctuations in occupancy factors for quantum levels that are given by the Poisson limit. Departures from Poisson statistic come from the fundamental symmetries associated with integer and half integer spin. Bose-Einstein symmetrization leads to enhanced occupation factors given by the Planck distribution (A=1,B=1) . Fermi-Dirac anti-symmetrization leads to reduced fluctuations which have A=1, B=-1 and are thus sub-poissonian. The disoriented chiral condensate has A=B=4/5 for neutral pions and A=B=1/5 for positive or negative pions [25,26]. A binomial distribution with mean Np has a variance Np(1-p) and therefore A=1-p and B=0.

It is also important to note that higher moments of the multiplicity distribution contain useful information that can help distinguish different distributions. For example, the skewness is defined by $<(N-<N>)^3>$ and is a higher moment that is often used. For a Poisson distribution the skewness is $<N>$. Factorial moments $F_q = <N(N-1)(N-2)…(N-q+1)>$ are also popular and were frequently used in the past when intermittency was a major concern. If $A \neq 1$, then a linear term exists in $F_2 = <N^2> - <N> = <N(N-1)> = (A-1)<N> + (B+1)<N>^2$. If $A = 1$, then $F_2 = (B+1)<N>^2$, which is quadratic in $<N>$. Thus a linear term in $<N>$ in the second factorial moment $F_2$ implies $A \neq 1$. The importance of factorial moments for high energy collisions can be found in several physics reports [27, 28, 29] where intermittency is also discussed.

## 3. Thermal Models

In this section results for thermal fluctuations will be presented. Thermal models give a very good description of single particle inclusive distributions in heavy ion collisions. The most recent results for relativistic heavy ion collisions can be found in refs.[8-10]. An early application of thermodynamic expressions to such collisions is to be found in ref.[30]. Statistical models based on the grand canonical partition function characterize mean particle distributions $N_j$ in terms of temperature T, freeze out volume V, and chemical potentials $\mu_j$

$$N_j = (1/2\pi^2) g_j TVm_j^2 \sum_{k=1}^{\infty} (\pm)^{(k+1)} K_2(k m_j/T) \exp(k \mu_j/T)/k \tag{2}$$

Here $g_j$ is the spin, isospin degeneracy of particle j. The + sign is for Bose-Einstein particles and the – sign is for Fermi-Dirac particles. The sum over k gives the degeneracy corrections with the k=1 term the Maxwell-Boltzmann limit. Equilibrium between all species limits the number of chemical potentials with this number determined by the number of constraints. For overall baryon number B and charge Q conservation two chemical potentials are needed, $\mu_b$ and $\mu_q$. Then $\mu_j = b_j \mu_b + q_j \mu_q$ where $b_j$ is the baryon number of particle j and $q_j$ is its charge. For symmetric systems of equal number of protons and neutrons, only one chemical potential is needed. The values of $\mu_b$ and $\mu_q$ are determined by $B = \sum b_j N_j$ and $Q = \sum q_j N_j$. The non-relativistic and non-degenerate limit of eq.(2) is

$$N_j = g_j (V/\lambda^3) \exp[k(\mu_j - m_j)/T] \tag{3}$$

The $\lambda$ is the thermal quantum wavelength $= h/(2\pi m_j T)^{1/2}$. For very small systems, a canonical ensemble description gives results that are significantly different than the grand canonical ensemble result. These difference lead to a canonical ensemble suppression factor [9].

## 3.1 Bose-Einstein and Fermi-Dirac degeneracy corrections and cycle lengths

As already noted, the Maxwell-Boltzmann limit has associated Poisson fluctuations. Corrections to the Maxwell-Boltzmann limit leads to a non-poissonian result. Here, we will study these changes. In previous papers [11,20] a physical interpretation was given to the sum over k that appears in eq.(2). This interpretation came from an approach based on Feynman's density matrix expansion in statistical mechanics. Specifically, symmetrization and anti-symmetrization in the density matrix lead to a cycle class decomposition of all the possible permutations. A given permutation can be decomposed into its cycle representation by specifying the number of cycles of length k in it. For example a particular permutation $1 \to 2 \to 4 \to 3 \to 1$ of 4 particles has one cycle of length 4. The identity permutation (no exchanges) of these 4 particles has 4 cycles of length 1. The permutation $1 \to 3 \to 1$ and $2 \to 4 \to 2$ has two cycles of length 2. The sum over k=1,2,3….in eq.(2) turns out to be the contributions of cycles of length 1,2, 3 …..Poisson statistics has only the first term or k=1 term present. Corrections to Poisson statistics come from k=2,3,4,..terms in eq.(2) which arise from cycles of length 2,3,4,… . Cycles of length two are the basis for HBT correlations. This cycle class approach can also be used to obtain a given particle probability distributions [11,21] and its associated mean value, fluctuation , skewness parameter, etc. Consider one type of particle with zero chemical potential as an example. Results for cases with $\mu \neq 0$ will be mentioned. A weight $W=W[n_1, n_2, …n_k, …]$ is assigned to any permutation, where $n_k$ is the number of cycles of length k. This weight is given by [20]

$$W = \Pi \ (x_k)^{n_k} / n_k ! \tag{4}$$

where the product is over all cycle lengths. The $x_k$'s that appear in this weight W are the individual terms of a given k that appear in eq.(2) divided by an extra k. Namely

$$x_k = (1/2\pi^2) g_s T V m^2 K_2(km/T)] / k^2 \tag{5}$$

This result applies to bosons and for fermions a factor of $(-1)^{(k+1)}$ is to included in eq.(5). For a particle with a chemical potential $\mu$, eq(2) is multiplied be $\exp(k\mu/T)$. The $g_s$ is the spin degeneracy and the subscript j is suppressed. The non-relativistic limit of eq.(5) is

$$x_k = g_s \ (V/\lambda^3) \exp(-km/T)] / k^{5/2} \tag{6}$$

This $x_k$ falls with k both exponentially in the mass and with a prefactor power law with exponent 5/2. The exponential factor introduces a length scale into the distribution of cycle length parameter through the presence of m/T. The importance of cycles of length 2 compared to cycles of length 1 is simply $(x_2/x_1)=\exp(-m/T)/2^{5/2}$. For m ~ T, this ratio is $1/(e 2^{5/2}) \sim 1/2^4$. If $\mu \neq 0$, the -m$\to \mu$-m. In the ultra-relativistic or zero mass limit,

the exponential Boltzmann suppression factor disappears and the exponent of the prefactor becomes 4 or

$$x_k = (1/\pi^2) g_s VT^3 / k^4 \tag{7}$$

The ratio $(x_2/x_1) = 1/2^4$ in the zero mass limit and this value is approximately the non-relativistic limit when $m \sim T$. The difference between the zero mass ultra relativistic limit and the non-relativistic limit is that the zero mass limit has no length scale, or is scale invariant in the cycle length weight factors. Cycles of all lengths contribute when taking moments of the $x_k$ distribution. This feature has important consequences for the multiplicity distribution as will be developed below. Scale invariance also appears near phase transitions where thermodynamic quantities can become singular. For example, the phenomenon of critical opalescence appears around a second order liquid gas phase transition where the scattering of light off droplets of all sizes produces this behavior. The distribution in size of these droplets falls purely as a power law and thus has no length scale. Similarly, the distribution of cycle lengths associated with Bose-Einstein condensation falls purely as a power law at the condensation point.

The probability distribution is the canonical partition function Z[A] to the grand canonical partition function, where Z[A] can be calculated recursively using $Z[A] = (1/A) \sum k x_k Z[A-k]$ with $Z[0]=1$. The sum runs over all cycle lengths from 1 to A. The grand canonical partition function is $Zgc = \exp(\sum x_k)$. The Zgc can be thought of as a generating function for the multiplicity distribution by introducing a quantity u and, in particular, $Zgc = \exp(\sum x_k u^k)$. This representation of Zgc shows that the $x_k$'s are the combinants of ref( ). Moreover the expansion of $Zgc = \sum Z[A] u^A$ relates Zgc to Z[A] as the $u^A$ term in this series, with $A=0,1,\ldots\infty$. The combinants are thus the cycle class weight factors, and thus they are given a physical interpretation here. Once the $x_k$'s are specified, moments of them give the mean, fluctuation, and skewness of the probability distribution. The connections are:

$$\begin{aligned}
\langle N \rangle &= \sum k x_k \\
\langle N^2 \rangle - \langle N \rangle^2 &= \sum k^2 x_k \\
\langle (N-\langle N \rangle)^3 \rangle &= \sum k^3 x_k
\end{aligned} \tag{8}$$

If only $x_1$ is present then the Poisson distribution follows. The $Zgc = \exp(x_1)$ and the $Z[N] = x_1^N/N!$ from the recurrence relation. From eq.(8), $\langle N \rangle = x_1$, so that the probability distribution is simply the result:

$$P_N = \langle N \rangle^N \exp(-\langle N \rangle) / N! \tag{9}$$

The fluctuation and skewness are also determined by the first term in the sums in eq.(8), so that the fluctuation is $\langle N \rangle$ as is the skewness. These features are all properties of the Poisson distribution. Including cycles of length 2 into the distribution results in departures from Poisson statistics. This is easily seen from the results of eq.(8) since the $<N> = x_1 + 2x_2$, while the fluctuation involves $x_1 + 4x_2$ and is no longer equal to $<N>$. Moreover, the presence of longer cycles with positive signs enhances the fluctuation above the unit cycle or Poisson limit. If we consider only cycles of 1 and 2, then the canonical partition function Z[N] is given by the following equation as can be obtained from the recurrence relation.

$$Z[N] = (1/N!) \sum (x_1)^{(N-2j)} (x_2)^j (N!/(j! (N-2j)!)) \tag{10}$$

The sum in eq.(10) runs from j=0 to [N/2] which is the integer part of N/2. When $x_2=0$ only the j=0 term is present in the sum which gives the unit cycle Boltzmann/Poisson limit. The Zgc for this case is $Zgc = \exp(x_1 + x_2)$ and the associated probability distribution for N particles is the ratio Z[N] / Zgc. Including longer cycles with + signs results in larger departures from Poisson statistics. For Fermions, cycles of length 2,4,… come in with negative signs which reduce the fluctuation with respect to the mean. The probability function can always be found by using the recurrence relation for Z[N] in all cases once the $x_k$'s are specified. The $x_k$'s can be related to the clan variables of Van Hove [14,15]. These variables were introduced to characterize departures from Poisson statistics, and where used to study properties of non-poissonian fluctuations seen in experimental data. Two clan variables Nc and $n_c$ are defined by Zgc=exp(Nc) and $n_c = <N>/Nc$. The Nc is the number of clans and $n_c$ the average number of particles in each clan. These variables can be simply related to the $x_k$'s :

$$Nc = \sum x_k \tag{11}$$
$$n_c = \sum k x_k / \sum x_k$$

For a Poisson distribution $Nc=x_1=<N>$ and $n_c=1$. If $n_c \neq 1$, departures from Poisson statistics are present. The Nc is also related to the void function $\chi = -\log(P_0)/<N>$ [13] where $P_0$ is the probability of finding no particle. Since $P_0 = Z[0]/Zgc = 1/Zgc$, the void function $\chi = 1/n_c$.

The zero mass limit for $x_k$ is given by equation eq.(7) has the following simple feature with respect to its moments given by the results of eq.(8).

$$<N> = (1/\pi^2) g_s V T^3 \zeta[3]$$
$$<N^2> - <N>^2 = (1/\pi^2) g_s V T^3 \zeta[2] = (\zeta[2]/\zeta[3]) <N> \tag{12}$$
$$<(N-<N>)^3> = (1/\pi^2) g_s V T^3 \zeta[1] = \infty$$

The zeta functions have values $\zeta[3] = 1.202$ and $\zeta[2] = 1.645$. The fluctuation is proportional to $<N>$ and is enhanced above it by a factor 1.369. The skewness diverges because of the presence of a pure power law fall off of the cycle lengths which fall as

$1/k^4$. This divergence in the third moment coming from the vanishing of the mass m can show up in thermodynamic quantities which will be discussed in sec(3.4). It should also be noted that the fluctuation result for zero mass particles falls into the representation of eq(1) with A given by the ratio of the two zeta functions shown in this equation and B = 0. In the non-relativistic limit of eq.(6), the Boltzmann factor exp(-k m/T) suppresses the contribution of very long cycles and all moments of the probability distribution are therefore finite. In the non-relativistic limit the fluctuation is not simply connected to the mean as in the zero mass limit. Rather, both involve a power series expansion in y = exp(-m/T). Defining $x = g_s V/\lambda^3$, $<N> = x \sum y^k/k^{3/2}$ and $<N^2> - <N>^2 = x \sum y^k/k^{1/2}$. These equations also apply to situations with a chemical potential, with $y=\exp((\mu - m)/T)$. Inverting the first of these series equations to find an equation for y in terms of $<N>/x$ and substituting this result into the second series for the fluctuation gives an expansion of the fluctuation in terms of a power series in $<N>/x$. This procedure results in:

$$<N^2> - <N>^2 = <N> + (<N>/x) <N>/2^{3/2} + (-1/4 + 2/3^{3/2})(<N>/x)^2 <N> \qquad (13)$$

If the above series is truncated at $<N>^2$ and the result is compared with the form of eq.(1), the coefficients A and B that appear in that equation can be identified. The A=1 and $B = 1/(x 2^{3/2}) = \lambda^3/(2^{3/2} g_s V)$. Coefficients involving $\lambda^3/V$ and powers of this ratio also appear in clusters formation and virial expansions. For clusters, these coefficients appear when the number of clusters of size 2,3,… is expressed in terms of nucleon numbers. In the case of statistical correlations, the coefficients are related to cycle lengths using the analogy that a cluster of size 2,3,…is similar to a cycle of length 2,3,… . In virial expansions of Fermi or Bose gases, these coefficients relate to the expansion of the pressure in powers of the density coming from quantum corrections to the ideal gas law. Such a virial expansion can be obtained by noting that the pressure equation is PV/T = $\sum x_k$. With $x_k = x y^k/k^{5/2}$ and $<N> = \sum k x_k$, a procedure similar to the one used to obtain the fluctuation in terms of a series in $<N>$ can be followed. Namely, inverting the power series for $<N>$ in terms of y to find y in terms of $<N>$ and now substituting into the pressure equation of state (partial pressure when other components are included), leads to the series:

$$PV/T = <N> - (<N>/x)<N>/2^{5/2} - (2/(9\sqrt{3}) - 1/8)(<N>/x)^2 <N> \qquad (14)$$

In eq.(12) the second term on the right hand side comes from cycles of length 2 which change the fluctuation from its Poisson value $<N>$, by the addition of a term involving $<N>^2$. In eq.(13), the second term on the right hand side also comes from cycles of length 2 which decrease the pressure for bosons because of the statistical "attraction" of bosons. For fermions, because of the extra factor $(-1)^{(k+1)}$ in $x_k$ the second term will be positive and the pressure will increase because of the statistical "repulsion" associated with the exclusion principle. . The skewness can also be developed as a power series in $<N>/x$ using the same technique and the resulting equation is:

$$<(N-<N>)^3> = <N> + (3/2^{3/2})(<N>/x)^2 <N> +(-3/4 + 8/(3^{3/2}))(<N>/x)^3 <N> \quad (15)$$

These power series expansions given by the last three equations for the fluctuation, pressure and skewness hold only in the non-relativistic limit. First order relativistic corrections can also be incorporated by using an asymptotic expansion for $K_2$ for large z leading to the following result for the $x_k$ :

$$x_k = g_s (V/\lambda^3) \exp(-km/T)(1+15T/(8km))/k^{5/2} \quad (16)$$

Since the exponential term is still present in this last equation of the cycle length weight, long cycles are suppressed and large fluctuations coming from statistics are not realized. The procedure of treating $y=\exp(-m/T)$ as an expansion parameter for inverting the power series can still be followed to express the fluctuation, pressure and skewness in terms of $<N>/x$. The coefficients in the resulting expressions for these quantities will now depend on (T/m). Defining a k-independent first order relativistic correction factor as $r=15T/(8m)$, the mean N expansion is now $<N>/x = y(1+r)+(1+r/2)y^2/2^{3/2}+...$ .Inverting this series to find y as a function of $<N>/x$, and then substituting the result into the corresponding series expansions for the fluctuation, skewness and pressure EOS gives the expansion of these quantities in powers of $<N>/x$. The new coefficients now depending on the first order relativistic correction factor r. Specifically, to second order in $<N>$, each of the second terms on the right hand side of the above equations is multiplied by $(1+r/2)/(1+r)^2$. For example, the second term for the fluctuation will now read $((1+r/2)/(1+r)^2)<N>(<N>/x)/2^{3/2}$. The first term is still $<N>$ in all these equations.

### 3.2 Role of decaying resonances and branching processes.

Resonance decays can lead to changes in fluctuations. This can be illustrated by the following simple example which is the decay of a neutral $\rho$ meson into two charged pions $\pi^+$ and $\pi^-$. Since each of these pions originates from a rho, the probability distribution is that of the rho. Also the number of $\rho$ = the number of $\pi^+$ = number of $\pi^-$ for this particular resonance decay. Since each charged pion has the probability distribution of the decaying rho, the fluctuation of either pion is that of its parent rho. If the rho probability distribution is Poisson, then its fluctuation is $<N(\rho)>$, where $N(\rho)$ is the number of $\rho$'s. The fluctuation in $N(\pi^+)$ is its mean number. By contrast, the total charge $N(ch) = N(\pi^+) + N(\pi^-) = 2 N(\rho)$ and the total charge fluctuation is $<N(ch)^2> - <N(ch)>^2 = 2<N(ch)>$, assuming a Poisson distribution $\rho$. A negative binomial NB distribution for $\rho$ or $<N(\rho)^2> - <N(\rho)>^2 = <N(\rho)> (1 + <N(\rho)>/a)$ leads to an NB distribution for either charged particle and a total charge distribution with variance = 2 $<N(ch)> (1 + <N(ch)>/(2a))$. In both cases, Poisson and NB for $\rho$, results in a non-Poisson or non-NB for the total charge, but a Poisson or NB for either charge distribution. Pions coming from several independently decaying sources, with each source having a Poisson distribution, gives a fluctuation that is the sum of the fluctuation of each. Let $N(\pi_1^+)$ and $N(\pi_2^+)$ be the number of positive pions coming from the decay of particle 1 and particle 2, each of which decays into one $\pi^+$ plus other particles. Then $<(N(\pi_1^+)+N(\pi_2^+))^2> = <(N(\pi_1^+))^2>+<(N(\pi_2^+))^2>+ 2<N(\pi_1^+)><N(\pi_2^+)>$ since 1 and 2 are independent so that the average of the cross term is the product of the average of each term. Consequently, the fluctuation for the total $\pi^+$ is the sum of the fluctuation of each. Moreover, since the fluctuation of each is a Poisson and therefore either $<N(\pi_1^+)>$ for particle 1 decay or $<N(\pi_2^+)>$ for particle 2 decay. Thus, the sum from each fluctuation produces the Poisson result of the sum or $<N(\pi_1^+)+N(\pi_2^+)>$. This simple result can be generalized to many decaying resonances. If a particular particle that decays has a non-poissonian fluctuation, then a $\pi^+$ from that decay has the same fluctuation as the original decaying particle and this is added with the fluctuations coming from other decays. The number of decaying particles of a particular type j is given by eq.(2).

A given resonance that decays may have several modes of decay specified by the various branching ratios to each of the channels. The following simple example will be used to illustrate the effect of a branching ratio on the multiplicity distribution. Let X be a particle that can decay into 2 channels with one channel containing a particle, called y, whose probability distribution is to obtained from that given by the parent particle X. The other channel does not contain this particle. The branching ratio to the channel with y is f and to the other channel this branching ratio is then 1-f. The situation considered in the previous paragraph is the f=1 limit of this more general situation. To proceed, the probability distribution of X is taken as Poisson. Then the mean of X is also its fluctuation. Let an event that produces the parent X particle have N such particles in that event. The number of y particles coming from the decay of X is determined by powers of the branching ratio f and can vary from 0 to N. The probability of having k particles of type y from this decay is the $y^k$ term in the binomial expansion of $(f+(1-f))^N$. Therefore the probability of having k particles of type y is given by the sum over the parent X Poisson distribution times this binomial factor. Specifically:

$$P_k(y) = \sum_{N=k}^{\infty} <X>^N \exp(-<X>)/N! \; \{f^k (1-f)^{(N-k)} N!/(k!(N-k)!)\} \tag{17}$$

The above sum can be simplified by writing $<X>^N = <X>^k <X>^{(N-k)}$ and noting that the sum $<X>^{(N-k)} (1-f)^{(N-k)} /(N-k)!$ which runs from N=k to $\infty$ gives $\exp\{-(1-f)<X>\}$. Consequently, the probability distribution of y is a Poisson with a mean $<y>=f<X>$ or:

$$P_k(y) = (f<X>)^k \exp(-f<X>)/k! \tag{18}$$

This result easily follows using $\exp(-<X>)\exp\{-(1-f)<X>\}=\exp(-f<X>)$ and also from simply combining the other terms. The variance or fluctuation of y is then its mean = $f<X>$.

Similarly, if the original X distribution were a negative binomial given by

$$P_N = ((N+a-1)!/(N!(a-1)!))(<X>/a)^N /(1+<X>/a)^{(N+a)} \tag{19}$$

then this distribution times the binomial distribution in the branching fraction f, summed as in eq.(17), regenerates a negative binomial distribution for the y particle that has <X> replaced by a mean <y> given by <y>=f<X>. The fluctuation in y is then

$$<y^2> - <y>^2 = <y>(1+<y>/a) = f<X>(1+f<X>/a) = f<X>+f^2<X>^2/a \qquad (20)$$

When the negative binomial parameter a=1, the resulting distribution is a geometric or Planck distribution. Therefore, these results also apply to a geometric or Planck distribution. Likewise, a binomial distribution for the parent particle X will generate a binomial for the branched particle y. If y is generated by a sequential processes from X as X → Z → y, with branching fractions $f_{XZ}$ and $f_{Zy}$, then y still carries the original distribution of X with <y>=$f_{XZ} f_{Zy}$<X>. This last result can also be generalized to an extended sequences of decays from X to y with various branching fractions, such as X → W → …. → Z → y. The original distribution is passed down to y generation by generation in this sequences from the original X particle to the final y particle, since each step in the sequence carries the same distribution of the previous step modified by a new mean.

If two resonances, X and Z, each independently contribute to y, with branching fractions $f_x$ and $f_z$, then the probability distribution for y is a convolution of each distribution, or

$$P_k(y) = \sum_{n=0}^{k} P_n(f_x<X>)P_{(k-n)}(f_z<Z>) \qquad (21)$$

The probability distributions on the right side of this equation are the original probability distribution for X and Z evaluated at the branching ratio modified means <X> and <Z>. If these original distributions are both independent Poisson distributions, the convoluted distribution is also a Poisson with mean <y>=$f_x$<X>+$f_z$<Z>:

$$P_k(y) = (f_x<X>+f_z<Z>)^k \exp(-f_x<X>-f_z<Z>)/k! \qquad (22)$$

Thus the variance is equal to <y> = $f_x$<X>+$f_z$<Z>. Also, the convolution of two independent negative binomials produces a distribution whose mean is the sum of the means of each negative binomial and whose variance is also the sum of the two variances. The variance would be $f_x$<X>(1+$f_x$<X>/$a_x$)+$f_z$<Z>(1+$f_z$<Z>/$a_z$). Similar remarks apply to two independent binomial distributions, or an independent Poisson and negative binomial. The convolution of a Poisson with mean <X> and a negative binomial with mean <Z> and parameter "a" produces a distribution which is

$$P_k(X, Z, a) = \{<X>^k \exp(-<X>)/k!\}(a<X>/<Z>)U[a, a+1+k, <X>(1+a/<Z>)] \qquad (23)$$

Here U is the confluent hypergeometric function and the term in curly brackets is the

Poisson distribution of <X> at probability number k. However, if there are correlations between any two distributions, and therefore their independence is no longer valid, then the above additive properties of the mean and variance are no longer true. The independence of the two distributions can be broken by correlations coming from constraints as will now be discussed.

### 3.3 Role of constraints in fluctuations and statistics

Constraints such as charge conservation in small systems can be very important in multiplicity distributions. For a small system with total charge 0, the number of positive charges = the number of negative charges. As an example consider baryon free matter and a system of charged pions. Then $N(+) = N(-)$, leaving out $\pi$ in the notation for simplicity. This result introduces a constraint in the canonical ensemble. Features of this constraint have been worked out in the Boltzmann limit [6,9]. Here, these results will be extended using the method developed in section 3.1 to include the role of statistics and also a net charge Q. First, the Boltzmann limit and this will then be developed further to include statistics using the cycle class representation of that section. The unconstrained partition function for both pions is

$$Zgc = \sum\sum (x_1^{N(+)}/N(+)!)(x_1^{N(-)}/N(-)!) = \exp\{2x_1\} \tag{24}$$

The $x_1$ is the given by eq.(5) with k=1. The unconstrained sums over $N(+)$ and $N(-)$ each run from 0 to ∞. When the charge conservation constraint is imposed, the Zgc, which we now call Zc, becomes a single sum over $N = N(+) = N(-)$ and is given by

$$Zc = \sum (x_1^2)^N/(N!)^2 = I_0(2x_1) \tag{25}$$

For the general case of net charge Q, so that $Q=N(+)-N(-)$, the two $x_1$'s in Zgc and Zc must be different. For simplicity of notation let $x_1 \to x$ for $N(+)$ and $x_1 \to y$ for $N(-)$. Then $Zc_Q = (x/y)^{(Q/2)}I_Q(2\sqrt{xy})$. For uncoupled Poisson distributions $<N(+)>-<N(-)> = Q = x-y$. This connection turns out to be also true for the canonical case. Consequently, the order of the Bessel function I must be consistent with the choice of x and y. In the charge constrained case, the

$$<N(\pm)> = \sqrt{xy}\, I_{Q\mp 1}(2\sqrt{xy}) \tag{26}$$

This can be obtained from the general result for the q factorial moment

$$<N(+)(N(+)-1)...(N(+)-q+1)> = (xy)^{q/2} I_{Q-q}(2\sqrt{xy})/I_Q(2\sqrt{xy}) \tag{27}$$

The factorial moment of the minus charge distribution has the order of the Bessel function in the numerator replaced by Q+q. Each successive term in the factorial moment $N(N-1)(N-2)...$ removes a + charge if $N=N(+)$ and thus reduces Q by 1, or increases Q by 1 if $N=N(-)$. If Q is initially negative then the order of the Bessel function for N(+) case is now $-Q-q$. and for the $N(-)$ case it is $-Q+q$. But $I_{-Q-q} = I_{Q+q}$ and $I_{-Q+q} = I_{+Q-q}$ so that the roles of the order of the Bessel function are reversed. This pattern of reducing the order of the canonical partition function when taking factorial moments has been used in the past [21]. The variance $<N^2>-<N>^2 = <N(N-1)>+<N>-<N>^2$, a useful expression when evaluating the fluctuation in N. Because of the constraint the fluctuation in $N(+)$ is equal to the fluctuation in $N(-)$, which can easily be obtained from $<Q^2>-<Q>^2 = 0$ and using $Q = N(+)-N(-)$. For $Q=0$, $x=y=x_1$, and for $q=2$

$$<N(+)> = x_1 I_1(2x_1)/I_0(2x_1) \qquad (28)$$
$$<N(+)^2>-<N(+)> = x_1^2 I_2(2x_1)/I_0(2x_1)$$

For small $x_1$ and thus small numbers of particles ($N \leq 3$), the canonical suppression factor is important. For $x_1 = 1$, which would have 1 particle on average in the unconstrained case, now has $<N(+)> = <N(-)> = 0.7$. As pointed out in ref.( ), the charge constraint plays an important role in the fluctuation. Specifically, the scaled variance or the variance divided by the mean, which would be 1 for a Poisson distribution, now goes to ½ at large $x_1$. Moreover, because of the strong correlation in $<N(+) N(-)> = <(N(+))^2> \neq (<N(+)>)^2$, the total charge fluctuation is $<(N(ch))^2>-<N(ch)>^2 = 4 (<(N(+))^2>-<N(+)>^2)$. Since the scaled variance of $N(+)$ goes to ½, the scaled variance of $N(ch)$ goes to 1. These results should be compared with resonance decays discussed in sect.(3.2). There, the scaled variance for a Poisson distribution of a neutral $\rho$ into two charged mesons gives a scaled variance of 1 for either meson and a scaled variance of 2 for the total charge. Thus, the fluctuations depend on the different sources – resonance decay or non-resonant decay but coupled by over-all charge conservation. The presence of a net charge Q can also modify the results. If x,y are $\gg 1$ and Q is small, the result is very close to the Q=0 result with the variances ½ the mean. When Q becomes a sizable fraction of x or y, the fluctuation is around ½ the sum of the means. Some examples are:
  1. x=625, y=576: Q=49, $<N(+)>$=624.75, $<N(-)>$=575.75, $\delta(N)$=300
  2. x=144, y= 64: Q=80, $<N(+)>$=143.79, $<N(-)>$= 63.79, $\delta(N)$=44.3
  3. x=144, y=121: Q=21, $<N(+)>$=142.67, $<N(-)>$=121.67, $\delta(N)$=65.9
  4. x= 64, y= 49: Q=15, $<N(+)>$= 63.75, $<N(-)>$= 48.75, $\delta(N)$=27.7
  5. x=  8, y=  6: Q= 2, $<N(+)>$=   7.75, $<N(-)>$=   5.75, $\delta(N)$= 3.

The above constrained results also apply to a charged resonance pair such $\rho^{\pm}$, where $\rho^{\pm} \to \pi^{\pm} + \pi^0$. Charge neutrality for a system of just $\rho$'s leads to a probability distribution for the $\rho^+$ or $\rho^-$ which is $(x_1^{N_+}/N_+!)^2/I_0(2x_+)$ neglecting symmetrization corrections to the Poisson limits.. With 100% branching into the above decay modes (f=1), the $\pi^{\pm}$ probability distribution of generated pions is also this distribution and the results found in the previous section apply. Note that the coupled $\rho^+$ and $\rho^-$ each contribute one $\pi^0$, so that the number of neutral pions is two per charged $\rho$ pair, but there is only one $\pi^+$ and one $\pi^-$ per $\rho$ pair. The neutral $\rho^0 \to \pi^= + \pi^+$ and also contributes to the yield of charged pions. Isospin conservation, charge conjugation parity or angular momentum coupled with statistics forbids $\rho^0 \to \pi^0 + \pi^0$.

The role of constraints including Bose-Einstein or Fermi-Dirac correlations can also be developed. These correlations can be incorporated into the previous Boltzmann limit by noting that the charged particle partition function for either N(+) or N(-) is changed from the unit cycle result $x_1^N/N!$ to $Z[N(+)]$ or $Z[N(-)]$ where the higher order cycles of length 2,3,...., can be included using the simple recurrence relation procedure developed in sect.(3.1). The result of eq.(10) shows how the partition function is changed when cycles of length 1 and 2 are only present. The unconstrained grand canonical partition function for both + and – pions is $\exp\{2x_1 + 2x_2\}$. When the constraint $N(+) = N(-)$ is imposed, eq.(25) with statistical correlations is $Zc = \Sigma (Z[N])^2$. To see how statistical correlations approximately modify the Boltzmann limit, only cycles of length 2 are included and results linear in $x_2$ are only kept. Then $Z[N] = x_1^N/N! + x_1^{(N-2)} x_2/(N-2)!$. The Zc is then $Zc = I_0(2x_1) + 2x_2 I_2(2x_1)$ and

$$<N(+)> = x_1(1+2x_2)I_1(2x_1)/\{I_0(2x_1) + 2x_2 I_2(2x_1)\} \tag{29}$$

The variance, to linear order in $x_2$, can be obtained from eq.(29) and the following result:

$$<N(+)^2> = \{x_1 I_0(2x_1) + 2x_2 (x_1 I_1(2x_1) + x_1^2 I_0(2x_1))/\{I_0(2x_1) + 2x_2 I_2(2x_1)\} \tag{30}$$

The factorial moment $f_2 = <N(+)(N(+)-1)>$ also follows from these two last results.

$$f_2 = \{x_1^2 I_0(2x_1)(1+2x_2) - x_1 I_1(2x_1)\}/\{I_0(2x_1) + 2x_2 I_2(2x_1)\} \tag{31}$$

When $x_2 = 0$, the resulting equation for the variance is that of ref.(6). The cycles of length 2 modify the results, but the main effect of the constraint is to reduce the scaled variance to a number very close to ½.

For heavy ion collisions, the initial charge and baryon number of the target and projectile are conserved. Strangeness is also produced with net strangeness = 0. The spectrum of particles (and anti-particles) used in the statistical models can be quite large [8]. For example, light non-strange mesons are: $\pi$, $\eta$, $\rho$, $\omega$, $\phi$, $\eta'$, strange mesons are: K, $K^*$, non-strange baryons are: p, n, $\Delta$, strange baryons are: $\Lambda$, $\Sigma$, $\Xi$, $\Omega$. Recursive methods in constrained canonical ensembles for particle production have been developed in ref.(11,21) and more extensively in ref.(31). For an ensemble of particles and anti-particles of these various types subject to the several global constraints listed, the canonical partition function for this system can easily be obtained by following a recursive procedure to obtain particle probability distributions. Results for this procedure will be given in a future publication.

### 3.4 Fluctuations and thermodynamic variables

In this subsection, thermodynamic variables that relate to fluctuations will be studied. The specific heat relates to energy fluctuations. For a system of pions or mesons with zero chemical potential contained in a volume V at a temperature T, the energy density E/V for a particular particle is given by

$$E/V = (1/2\pi^2) g_s g_I m^3 T \sum_{1}^{\infty} \{(3/4)K_3(km/T) + (1/4)K_1(km/T)\}/k \tag{32}$$

The $g_s$ and $g_I$ are the spin and isospin degeneracy factors with $g_s=1$ and $g_I=3$ for pions. The sum over from k=1 to ∞ is over the cycle lengths. The derivative of E with T and V held fixed is the heat capacity at constant V. Let $C_v = \partial(E/V)/\partial T$ and allow m = m(T), then

$$(2\pi^2) C_v / g_s g_I = 3m^3 \sum_{1}^{\infty} K_3(km/T)/k + m^2(dm/dT)T \sum_{1}^{\infty} K_1(km/T)/k +$$

$$(m^4/T - m^3 dm/dT) \sum_{1}^{\infty} K_2(km/T) \tag{33}$$

In the non-degenerate limit, only the k=1 term is present. In the ultra relativistic limit as m → 0, then

$$C_v = (1/\pi^2) g_s g_I \{12T^3 \zeta(4) - m(dm/dT)T^2 \zeta(2)/2\} \tag{34}$$

The $\zeta$ functions come from sums over cycle lengths from k=1 to k=∞. For situations were the mass m independent of T the last term on the right hand side of this equation is zero and the first term is the Stefan-Boltzmann limit for a zero mass particle. If a particle has a temperature dependent mass then the dm/dT terms contribute in either eq.(33) or its ultra relativistic limit given by eq.(34). Moreover, if a particles mass approaches zero at some critical temperature Tc as $m(T) = m_o(1-T/Tc)^\gamma$, then the second term in the last equation has $m(dm/dT) = -\gamma m_0^2 (1-T/Tc)^{(2\gamma-1)}/Tc$ which can go to zero if $\gamma > ½$, go to a constant if $\gamma = ½$ or diverge if $\gamma < ½$. Higher order derivatives of E with respect to T can be easily obtained. Thus temperature dependences in the mass of particles can increase the specific heat $C_v$ and consequently the energy fluctuations in the system depending on the coefficient $\gamma$. If the temperature dependence of the mass is neglected, then $C_v/<N> = 12\zeta(4)/\zeta(3)$ in the ultra relativistic limit of an ideal Bose gas of massless particles. Resonance decays also contribute to $C_v$. The non-degenerate (k=1 term) or Poisson limit for a particle mass m >> T has

$$C_v = g_s g_I (m^2/T^2) \exp(-m/T)(1+39T/8m)/\lambda^3 \tag{35}$$

with a particle density

$$N/V = g_s g_I \exp(-m/T)(1+15T/8m)/\lambda^3 \tag{36}$$

The ratio $C_v/(N/V) = (m/T)^2 (1+39T/8m)/(1+15T/8m)$ in this non-degenerate $m \gg T$ limit. The $\rho$ meson has an spin-isospin degeneracy factor of 9 which enhances its contribution to $C_v$ even though the mass of $\rho$ is 775 MeV which is 4-6 times a typical temperature that can be reached at RHIC. Taking $m/T = 4$, the ratio of contributions of $\rho$ to $C_v$ compared to $\pi$ is about 1 when the pion's contribution is taken in its massless limit. Each $\rho$ contributes 2 pions and the density of $\rho$'s compared to $\pi$'s is .38 in this limit of massless pions. This ratio of densities is .45 when the mass of the pion is taken to be 137 MeV, its value in free space at this temperature of 775/4 MeV. At lower T's the ratio of these densities is reduced considerably.

Fluctuations also play an important role in the hadronic compressibility. Density fluctuations are connected to the isothermal compressibility $\kappa = -(1/V)(\partial V/\partial P)_T$ through the result $\langle N^2 \rangle - \langle N \rangle^2 = \langle N \rangle^2 T\kappa/V$. In fact, in ref.(32), event-by-event fluctuations are proposed as a method of determining the hadronic matter compressibility using this relation. In this reference, this connection is generalized to a multicomponent system with the result

$$\sum_{i=1}^{m} \langle N_i \rangle^2 /(\langle N_i^2 \rangle - \langle N_i \rangle^2) = -(V^2/T)(\partial P/\partial V)_{T,\{N_i\}} \tag{37}$$

In a one-component system of massless particles with zero chemical potential the left hand side of this equation is $\varsigma(3)(6/\pi^2)\langle N \rangle$, where $\varsigma(2) = \pi^2/6$ [32]. When the form of eq.(1) applies for the fluctuation, then the left hand side is modified to $\langle N \rangle/(A(1+B\langle N \rangle))$. Since the Poisson limit has A=1, B=0, the left hand side is simply $\langle N \rangle$ in this limit.

An incompressibility coefficient is often used in nuclear physics which has to do with the curvature of the energy with density and is defined by $K = 9V^2 \partial^2(E/A)/(\partial V)^2$, with T and $\{N_i\}$ held constant, and evaluated at the minimum E/A. At T=0, the $K \approx 200$ Mev with a minimum nuclear matter density. At higher T a minimum exists at higher density and the value of the incompressibility can become much larger[33], reaching values of 1000MeV. This dramatic increase is associated with the appearance of strong nucleon-nucleon correlations.

## 4. Conclusions

This paper centered around a study of the role of fluctuations in the statistical model of relativistic heavy ion collisions. The statistical model for particle production assumes equilibrium is developed in the collision of two heavy ions. Equilibrium results when the particle production rates are equal to the particle annihilation rates. This must happen within the time scale of the reaction or within an associated expansion. If equilibrium is established then simple expressions, independent of the production mechanism and associated cross sections, give particle multiplicity distributions. These multiplicity distributions depend only on a few parameters related to the temperature of the system which assumes thermal equilibrium, volume of the system where equilibrium is established, and chemical potentials for conserved quantities which assume chemical equilibrium. Equilibrium also destroys any persistence of memory of prior states in the space time evolution of the system.

   The statistical model has been successful in accounting for the mean particle multiplicities. Little has been done in the statistical framework with respect to the actual particle probability distributions and their associated higher moments. The higher moments involve the fluctuations, skewness and factorial moments. Event-by-event studies, which are gaining considerable attention, carry such information. Large non-poissonian fluctuations have been reported in the multiplicity distributions generated in the collision of small systems such as the H1 and L3 experiments. Such fluctuations can be used to distinguish between different models. Various aspects of the event-by-event structure of the statistical model of heavy ion collisions are developed in this paper. The role of statistics, relativity, constraints, decaying resonances and branching processes are studied in this framework. A simple method of including statistics into the statistical distributions is developed in terms of cycle lengths associated with permutations that arise in the density matrix. This matrix has to be made totally symmetric for bosons or anti-symmetric for fermions. In the non-relativistic limit small cycle lengths dominate because of Boltzmann factor in mass over temperature which introduces a scale into the problem. In the zero mass limit, distributions become scale invariant. Consequences associated with this scale invariance are discussed, such as enhanced fluctuations and divergences in the skewness of the probability distribution of a particle. A particle that originates from the decay of another particle with some branching fraction to produce that particle is shown to carry the same probability distribution as the original parent particle with a mean that is modified by the branching fraction. This is also shown for a sequence of associated decays which eventually end in the particle of interest. How to incorporate both constraints and statistics in a simple way is also discussed in terms of the cycle class representation. The relation between fluctuations and thermodynamic properties is developed for the specific heat. The specific heat is shown to be sensitive to the temperature dependence of the masses of the particles. A vanishing mass at some critical temperature can lead to a divergence of the specific heat for some values of a critical exponent.

## Acknowledgements
The author would like to thank the Department of Energy for a grant (grant number DE-FG02-96ER-40987) which supported in part this research.